# Magnetic domain and magnetic resistance phase transition in strongly correlated electronic material of perovskites junction

Ren R†, Weiren Wang, Xuan Li, Zhongxia Zhao

*Department of optics , Xian Jiao Tong University, Xi'an, 710054, China*

The junction magnetoresistivity and domain phase transition were studied between ZnO and $La_{0.4}Gd_{0.1}Sr_{0.5}CoO_3$ thin films grown on $LaAlO_3$ (100) substrates epitaxially by pulse laser deposit. The ferromagnetic transformation into phase-separated (two phase) state was displayed below Tc~127 and has observed that the lattice change discontinuously in the doped cobalt perovskites $La_{0.4}Gd_{0.1}Sr_{0.5}CoO_3$. The Ginzburg-Landau phase field is introduced to deduce antiferroelectric domain structure in LGSCO thin film. On the basis of the domain structures, the phase boundary of thin film is strongly dependent on the combination of electric-mechanical coupling. The phase transformation into phase separated state occurs below Tc~127-128K, and have displayed that the lattice constants change discontinuously at the transformation. The positive MR of ZnO/LGSCO heterojunction exhibited the MIT behavior at 0.2 T is 4.86%, at 0.5 T is 6.05% for approximately 140K.

Author to whom correspondence should be addressed mail: renrnano@126.com

## 1. INTRODUCTION

The phase boundary effect in doped perovskite oxides have attracted much attention of material science since coupled ferromagnetic ordering and insulate-metal transition, as well as the colossal magnetoresistance CMR phenomenon which make $R_{1-x}A_xCoO_3$ $ZnO/La_{0.4}Gd_{0.1}Sr_{0.5}CoO_3$ heterojunction an interesting candidate for ferromagnetic and ferroelectric applications. Especially, CMR effect is strengthened near the phase separated between the charge ordered insulator and the ferromagnetic metal. $R_{1-x}A_xCoO_3$ has a rhombohedral perovskite structure, lead to altered spin ordering and phase boundary, which is due to $Co^{3+}$ and $Co^{4+}$ having spin $Co^{3+}$ $t^6_{2g}e^0_g$, $t^5_{2g}e^1_g$, and $t^4_{2g}e^2_g$ under various doped concentration and temperature. [1,2]. ZnO is generally regarded as n-type due to oxygen impurity, and their carrier concentration is dominated by oxygen deficiency. So far, many theoretical and experimental studies have been carried out on the phase boundary problem. Among them, LSCO, which is ferromagnetic but is transferred in to antiferromagnetic charge insulator (CI) phase below $T_{CI}$ have exhibited conexistence of FM and CI microdomains.[6] Phase separatation in rare-earth metals of hydrogen-concentration have been investigated at ambient temperature and high pressure by x-ray diffraction changed in structural, electronic, and magneticproperties.[7] Goodenough reported the structural and magnetic properties of the metallic double-perovskite system, and suggested that ferromagnetic long-range ordered domains are coupled antiferromagnetically across antiphase boundaries, random disorder within domains may be small.[20],[27]Yacaman et al. investigate the self-assembly of this material through controlling the material structure and chemical order and found that bimetallic core/shell has the structure of faced centered cubic octahedron. [19],[22] The FM metallic state and the charge/orbital ordered insulator Cr-doped $Nd_{0.5}Ca_{0.5}MnO_3$ have ferromagnetic (FM) microdomains in the phase-separated state and FM microdomains.[9] Rivadulla studied perovskite oxide spin glass phase separation behavior. [18]

Recently, it is found that $LaCoO_3$ was charge transfer insulator,[9],[10] while $La_{0.7}Sr_{0.3}CoO_3$ was intermediate between charge transfer and Mott Hubbard-type compounds.[11] The spin ordering, charge, and orbit ordering transition observed at around the half-filling is strongly coupled with lattice distortion. Chen explored the electronic transport of $Al:ZnO/BiFeO_3/ITO$ junction related to electron spin-dependent scattering and the interface resistance. [16] In the DE mechanism, the ferromagnrtic phase transition was suggested to be introduced by hopping of eg electrons between $Co^{3+}$ and $Co^{4+}$ through the oxygen ions that also result from conductivity in carriers transport. The heterojunction quantum energy band was modified by spin, orbital ordering, and lattice field to exhibit photoinduced resistance and colossal magnetoresistance (CMR) effect. [1,2,4,7] However, the phase transition and domain separation mechanism of ZnO/LGSCO heterojuction are little explored from the direction of junction CMR effect in the heterostructure in the past few years. Furthermore, the $La_{0.4}Gd_{0.1}Sr_{0.5}CoO_3$ topological evolutions by 2D grain growth using a continued diffuse interface field are little studied. The ferromagnetic and charge properties are attributed to the further understanding of charge ordering, magnetic order, electric-mechanical coupling, spin-orbital coupling, optical field and lattice structure.

The domain wall orientations structure and spin-orbital coupling strongly influence the ferroelectric and antiferromagnetic properties of $ZnO/La_{0.4}Gd_{0.1}Sr_{0.5}CoO_3$. The LGSCO material was accompanied from ferromagnetic to antiferromagnetic transition, which is called double exchanged P doping. $La_{0.4}Gd_{0.1}Sr_{0.5}CoO_3$ from ferromagnetic to antiferromagnetic ordering was accompanied by a $Co^{3+}$ / $Co^{4+}$ charge orbit ordering. The mechanics and electric coupling of $ZnO/La_{0.4}Gd_{0.1}Sr_{0.5}CoO_3$ thin films strongly influence domain structure and domain wall. Domain walls can produce pinning sites that reduce remnant polarization, as well as domain structure plays a role in ferroelectric and ferromagnetic order.

In this work, we present the structural analysis elucidated the phase separated state in the heterojunction of cobalt perovskites $La_{0.4}Gd_{0.1}Sr_{0.5}CoO_3$ and ZnO films fabricated by pulse laser deposition consist of ZnO and $La_{0.4}Gd_{0.1}Sr_{0.5}CoO_3$ grown on LaAlO substrate. The IMT effect is enhanced near the phase boundary between the ferromagnetic metallic(FM) and the charge-ordered insulating phase, where the phase separation effect is also enhanced. The $La_{0.4}Gd_{0.1}Sr_{0.5}CoO_3$/ZnO junction exhibited excellent rectifying behavior and MR properties over the temperature range of 80-300K. The Ginzburg-Landau phase field is used to evolution antiferroelectric domain structure LGSCO thin film. The temperature dependence of mechanics and electric coupling properties determined the phase properties of LaGdCoO and ZnO thin films. The phase-field of domain structure in epitaxial ZnO/$La_{0.4}Gd_{0.1}Sr_{0.5}CoO_3$ and gain size effect of our sample have been investigated.

## 2. EXPERIMENTS

In order to obtained an appropriate chemical composition for present study, the target $La_{0.4}Gd_{0.1}Sr_{0.5}CoO_3$ was prepared from the analytically pure oxides used proportion of $La_2o_3$, $Gd_2O_3$, $Co_3O_4$ and $SrCO_3$ by solid state reaction, after repeated grinding and sintering at 1250 °C for 24h for $La_{0.4}Gd_{0.1}Sr_{0.5}CoO_3$. The ZnO was sintering at 500°C for 48h by solid state reaction. The LGSCO and ZnO layers were successively deposited on the single-crystal LAO (100) substrate by pulse laser deposition. The multilayer thin films were annealed at 800°C with an oxygen pressure 6 Pa for 5h in order to get better epitaxial character and oxygen doped deposit. The structure of the target and the orientation of the deposited film were studied by XRD (Bruker D8 Advance XRD, 40KV, 40mA, smallest angular step $0.0001^0$). The sample of heterojunction thin film was placed in a JanisCCS-300 closed-circuit refrigerator cryostat (JanisCCS-300) over the range from 50 to 300 K. The magnetic field of 0.2 and 0.5 T was supplied by the electromagnet perpendicularly applied to heterostructure.

The MR and ρ-T in film was studied in a JanisCCS-300 over the range from 50 to 300 K at 0.2 and 0.5T, respectively. The topography of ZnO/$La_{0.4}Gd_{0.1}Sr_{0.5}CoO_3$/LAO was measured though the AFM image. The AFM surface morphology and phase field image for ZnO/$La_{0.4}Gd_{0.1}Sr_{0.5}CoO_3$/LAO and ZnO/ $La_{0.4}Gd_{0.1}Sr_{0.5}CoO_3$/Si were charactered by atomic force microscopy (AFM), respectively.

## 3. RESULTS AND DISCUSSION

FIG.1 a, b show the AFM phase field image of the surface morphology ZnO/ $La_{0.4}Gd_{0.1}Sr_{0.5}CoO_3$/LAO. The bright region was hard phase, and the dark region was soft phase. The FIG. show the surface roughness for ZnO/ $La_{0.4}Gd_{0.1}Sr_{0.5}CoO_3$/Si was 2nm and the surface quality is very well.AFM results were presented for ZnO/$La_{0.4}Gd_{0.1}Sr_{0.5}CoO_3$/LaAlO and ZnO/ $La_{0.4}Gd_{0.1}Sr_{0.5}CoO_3$/Si,respectively. ZnO/$La_{0.4}Gd_{0.1}Sr_{0.5}CoO_3$ film displayed very smooth surface with uniform grain size of 31nm and the interface roughness of ZnO/$La_{0.4}Gd_{0.1}Sr_{0.5}CoO_3$ was about 2.5nm due to LaAlO substrate.

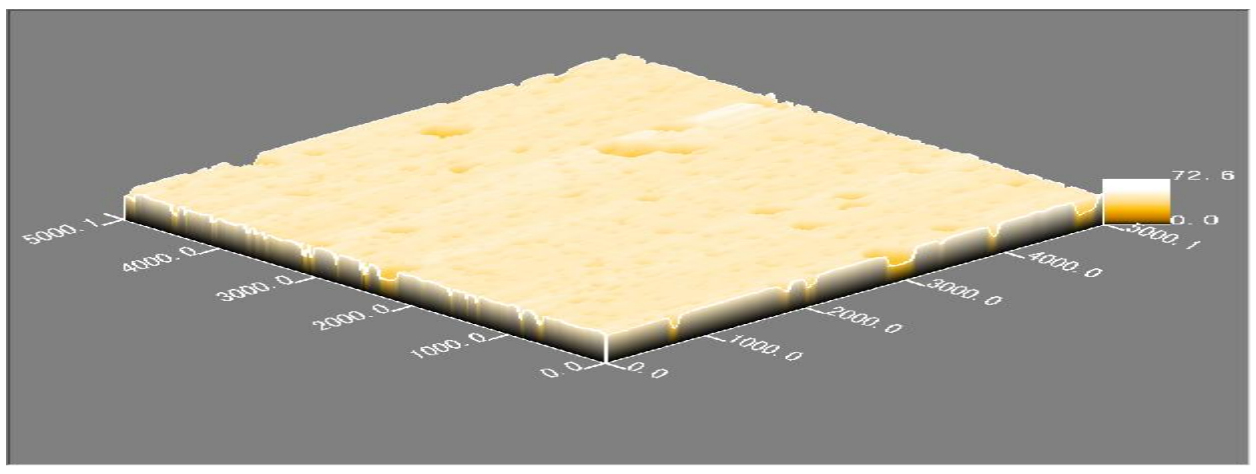

(a)

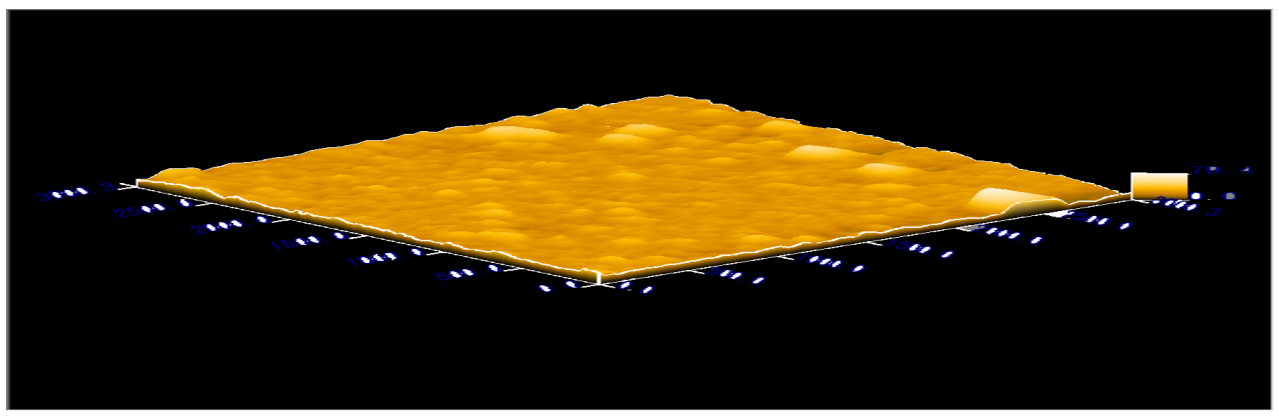

(b)

FIG.1. a. AFM phase field image of the surface morphology ZnO/ $La_{0.4}Gd_{0.1}Sr_{0.5}CoO_3$/LAO. The bright region is hard phase，and the dark region is soft phase. FIG.1.b The AFM image of surface morphology ZnO/ $La_{0.4}Gd_{0.1}Sr_{0.5}CoO_3$/LAO.

Fig.2 shows an evidences of two phase Rietveld refinement at 113 K(<$T_{CO}$). These two perovskite phases can be characterized by the length of c, long c and short c respectively. The final refinement is satisfied with R3c group. We explored that the crystal symmetry group at room temperature is orthorhombic (Pbnm) without detected impurities. Fig.2(a) showed the sample of the Rietveld refinement at 240 K. Fig.2(b) indicated that the XRD are satisfactory, these two perovskite phases as result of single orthorhombically disordered perovskite (Pbnm;) and rhombohedral ( $R\bar{3}c$ ).

The LGSCO/ZnO structure was characterized by x-ray diffraction pattern at 220K shown in Fig. 2. Two crystalline phases were identified from the XRD patterns. One was ZnO film crystalline phase of a hexagonal wurtzite. Other was LGSCO crystal phase of hexagonal perovskite structures (space group:R3c) with no more than 1% $Co_3O_4$ as the second phase by grain refinements.

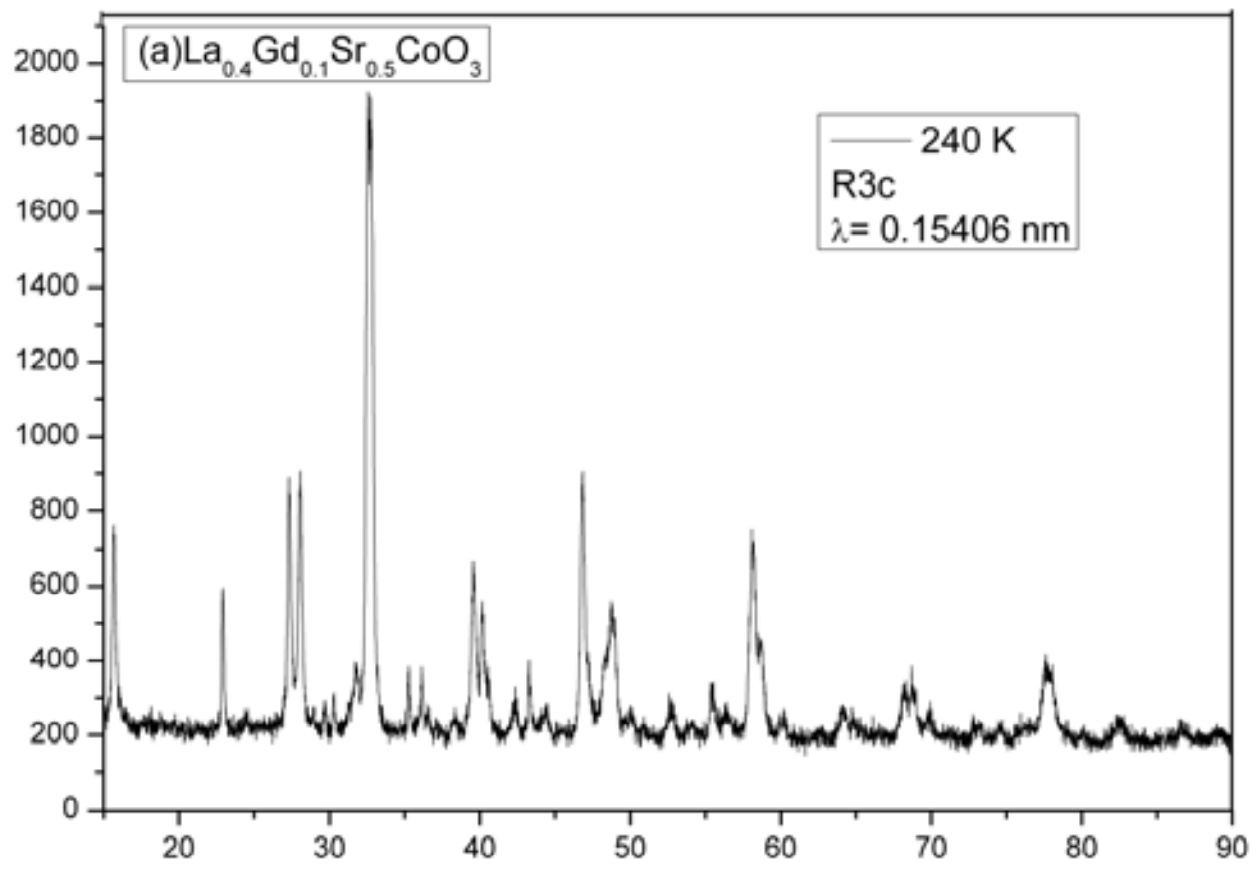

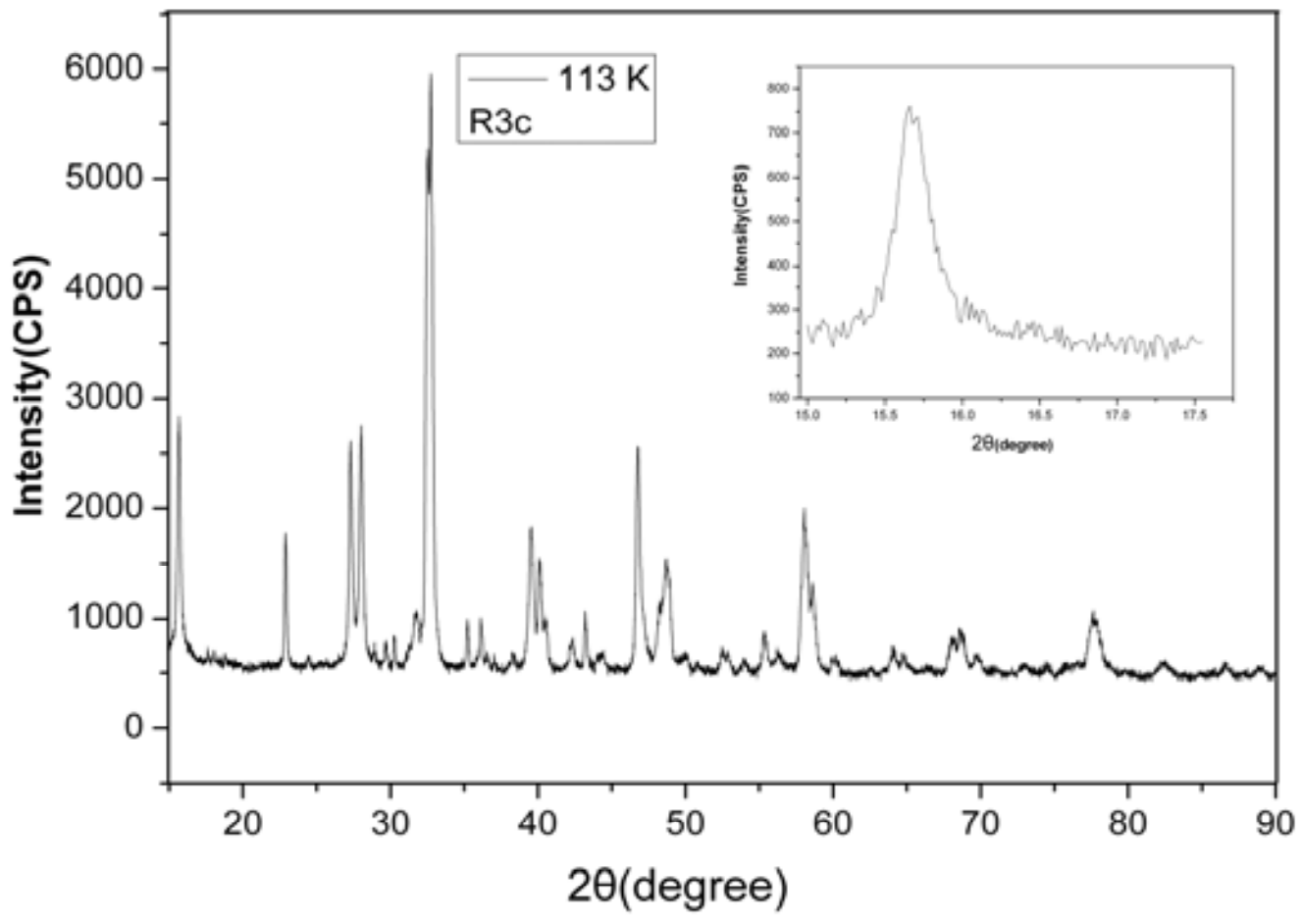

(b)

Fig.2 X-ray powder pattern for $La_{0.4}Gd_{0.1}Sr_{0.5}CoO_3$ (a) at 240 K and (b) at 113 K. Solid curve is the result of the Rietveld analysis with (a) single orthorhombically distorded perovskite (Pbnm;) phase from $La_{0.5}Sr_{0.5}CoO_3$ rhombohedral ( $R\bar{3}c$ ) to $Gd_{0.5}Sr_{0.5}CoO_3$ orthorhombic (Pmmm), and with two perovskite phases, respectively.

It is known that the charge and orbital ordering transition at Gd half filling(x~0.5) is strongly coupled with lattice distortion. The $CoO_6$ is distortion of oxygen octahedral. The LGSCO lattice parameters, Co-O bond length and Co-O-Co bond angle, were adjusted by $Gd^{3+}$ ions size, and dependence on average radio of A (La, Gd) site ion <A>. The Fig.3(a) image indicated that the $La_{0.4}Gd_{0.1}Sr_{0.5}CoO_3$ (100) and (200) diffraction peaks occurred at $2\theta=32.8^0$ and $2\theta=47.2^0$, while the Fig.3(b)ZnO (100) (002) (101) (102) (110) (103) (200) diffraction peak occurred at $2\theta=31.59^0$,$2\theta=34.37^0$,

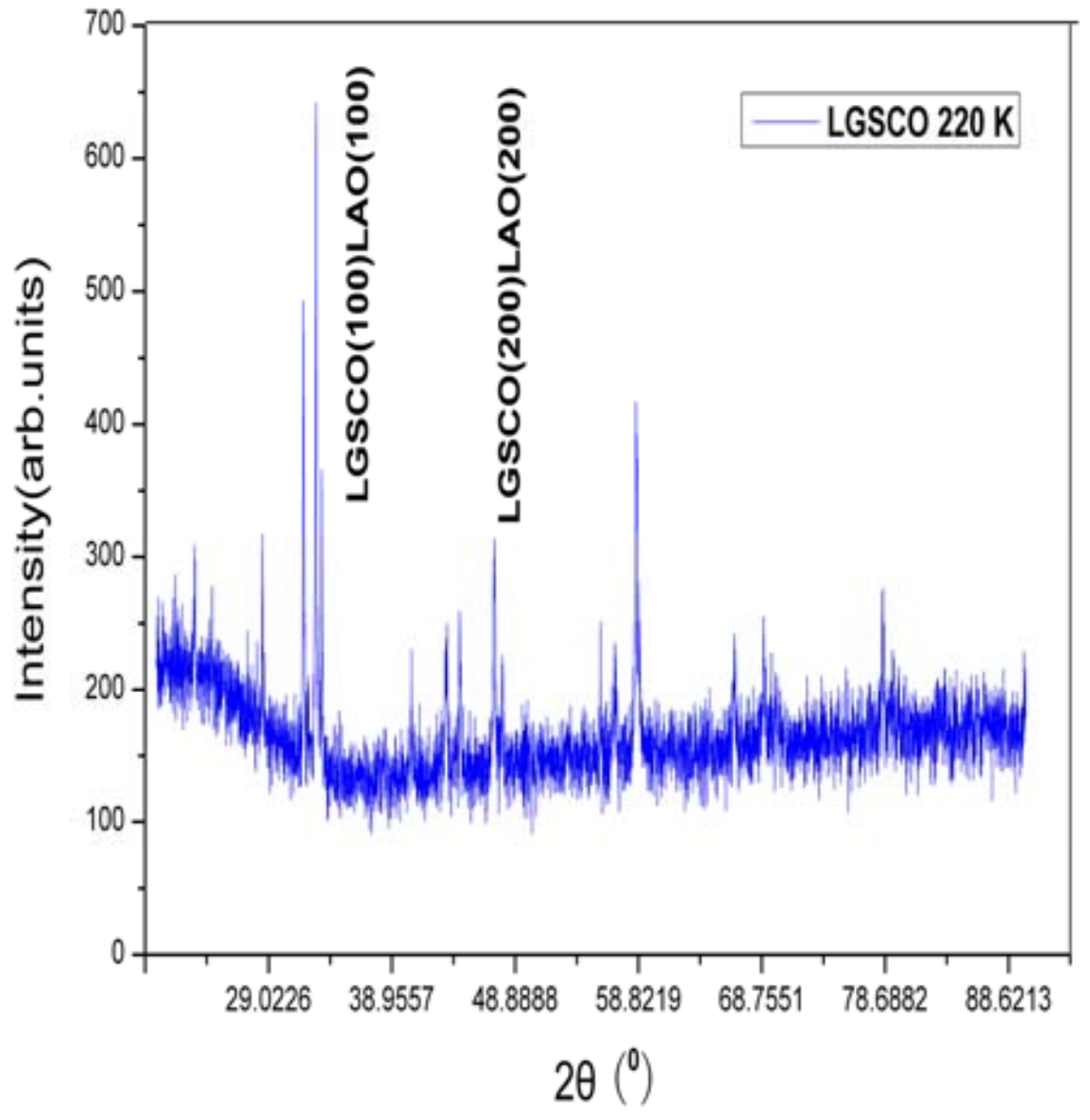

(a)

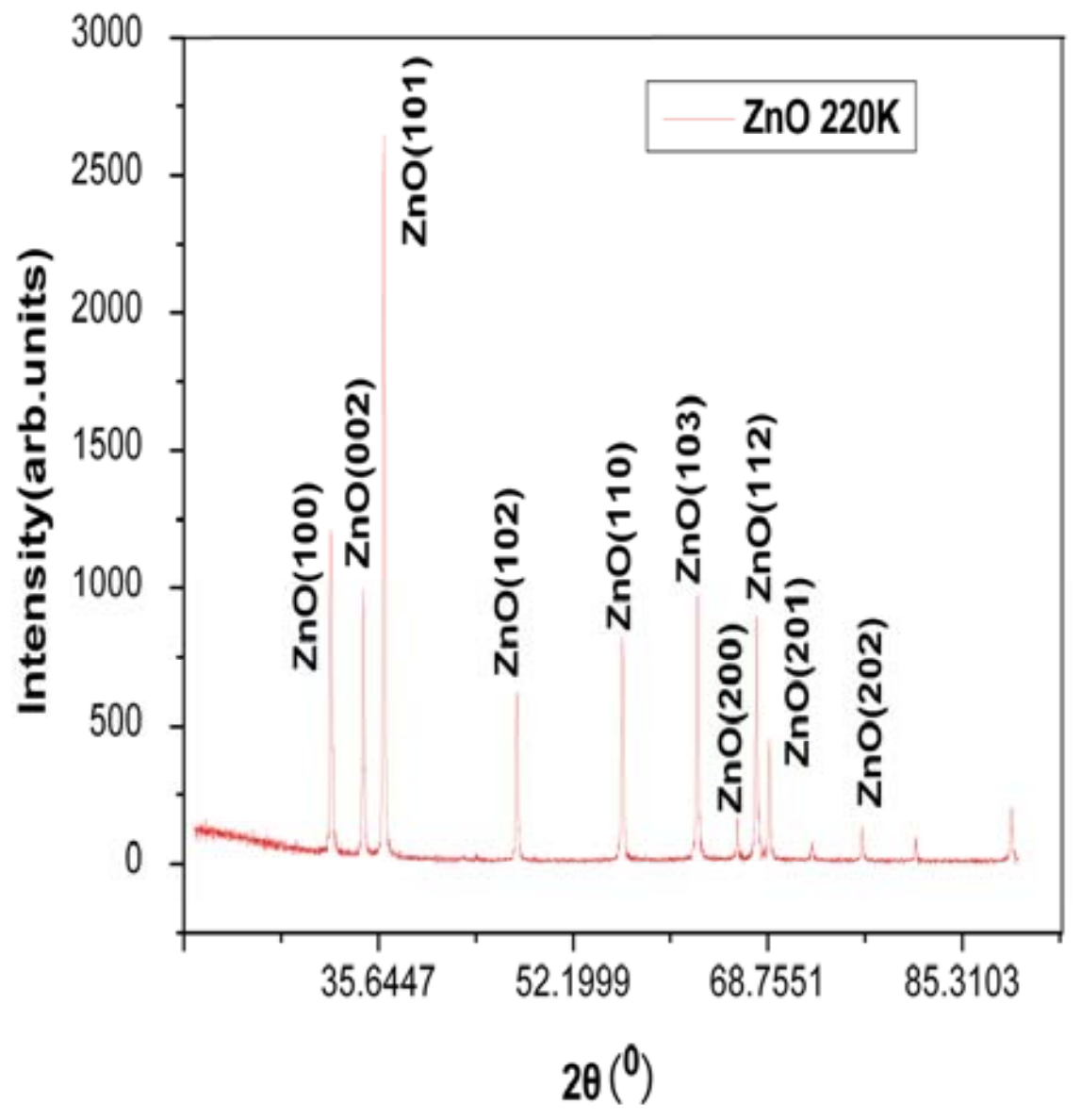

(b)

Fig.3 X-ray diffraction of ZnO/$La_{0.4}Gd_{0.1}Sr_{0.5}CoO_3$/LAO heterostructure at 220K. The Figure (a) (b) shows the (100) and (200) peaks of the LGSCO film (a) and ZnO film (b).

$2\theta=36.09^0$, $2\theta=47.3^0$,$2\theta=56.4^0$, $2\theta=62.78^0$and $2\theta=66.15^0$. The spectrum peak showed crystal indices of LGSCO. The diffraction peaks of LAO (100), (200) substrate appear at $2\theta=23^0$and $2\theta=47^0$. This chart identified LGSCO film in the crystal plane (100) identically matching with the LAO substrate with multi-crystal structure. The XRD of $La_{0.4}Gd_{0.1}Sr_{0.5}CoO_3$ has GdO characteristic diffraction peak at $2\theta=28^0$. The doped Gd goes in to the LGSCO lattice. Because the $Gd^{3+}$ electronic structure is $4f^7$, $Gd^{3+}$ ion radius is smaller than $La^{3+}$ and $Sr^{2+}$. The $Gd^{3+}$ has an effect on both structure factor and electronic state balance.

Table 1, structure and properties of $La_{0.5}Sr_{0.5}CoO_3$ and ZnO ($\mathring{A}$)

| Ln | A | ZnO | Space group | $<r_A>$ | $\sigma(\mathring{A}^2)<\mathring{A}>$ | Lattice parameter ($\mathring{A}$) |
|---|---|---|---|---|---|---|
| | | | | | | a b c |
| La | Sr | | R3c | 1.400 | 0.0016 | 5.4152 |
| Gd | Sr | | Pnma | 1.329 | 0.0123 | 5.3746, 7.5601,5.3723 |
| $La_{0.255}Gd_{0.245}Sr_{0.5}CoO_3$ | | | R3ch | 1.363 | 0.0081 | 5.3948 |
| $La_{0.4}Gd_{0.1}Sr_{0.5}CoO_3$ | | | Pm3m | 1.329 | 0.00468 | 3.8402 |
| ZnO | | | P63mc | 1.99 | 0.0471 | 3.2511, ,5.1993 |

We listed the weighted average radius given in Table 1.[30] The $La_{0.5}Sr_{0.5}CoO_3$ have representative 12 coordination number of A sited cations, as well as $Gd_{0.5}Sr_{0.5}CoO_3$ representative 9. The structure exhibit that when A=Sr, the $LnACoO_3$ structure is rhombhohedral (space group: $R\bar{3}c$) until Ln =La. The Sr-site coordination number is 12 when the crystal structure is rhombhohedral, and 9 when it is orthorhombic. In the situation of A=Sr, Ln=La, Gd, we could get an satisfactory good fit for the Pnma, Pm3m and $R\bar{3}c$ space groups, and we have, furthermore, preferred the space group with higher symmetry as per the normal practice. The rhombhohedral and orthorhombic structure illustrate how the $CoO_6$ octahedra is distorted in the orthorhombic structure, especially in $R\bar{3}c$ and Pm3m space group of the $La_{0.4}Gd_{0.1}Sr_{0.5}CoO_3$. The Co-O bond increase with the variation from $La_{0.5}Sr_{0.5}CoO_3$ to $La_{0.4}Gd_{0.1}Sr_{0.5}CoO_3$, at the same time the Co-O bond decrease.

The antiferromagnetic spin-ordering accompanied a charge-order transition with CO phase structure exhibited a strong correlated coupling between the spin and orbital degrees of freedom. The diffraction peak of $ZnO/La_{0.4}Gd_{0.1}Sr_{0.5}CoO_3$ was matched with the LAO substrate (lattice constant 0.38769 nm) when the lattice constant was 0.386 and 0.379nm respectively. The results indicated that the $La_{0.4}Gd_{0.1}Sr_{0.5}CoO_3$ and ZnO thin films had better epitaxial characters respectively. The ZnO film is $2\theta=34^0$ for the (002) orientation with the structure of a hexagonal wurtzite. According to the width of half peak, XRD wavelength, diffraction angle and proportional constant K, we calculated the average grain size of $LaGdSrCoO_3$ film using the Scherrer equation. The $LaGdSrCoO_3$ average grain size of the film was 30.7 nm in coincidence with AFM results.

**(1). Electronic feature of phase separated state**

Figure 4, we showed the temperature dependence of the ZnO/LGSCO junction resistance, and Figure 5 MR of compositions of LGSCO at variation 0.2T, 0.5T. The magnetoresistance mechanism in perovskite cobalt oxides attracted special interested due to its Tc sensitivity and cation-size mismatch applications. The most important message of Fig.2 is that domain boundary and lattice mismatch. The lattice constant shows a discontinuous change at Tco. In another word, the system is transformed into phase separation. Such a paramagnetic state at Tco 200 K is described to the random nucleation of a low temperature phase and subsequent stress-induced growth of the secondary phase. The phase separation is considered stress-induced phase separation which is origin from lattice constant. At Tco 200 K, a b and the short c is for paramagnetic insulator while those ferromagnetic is for the long c phase. To determine critical temperature, i.e., Curie temperature Tc and the charge-ordering temperature Tco, we have tested temperature dependence of resistivity and magnetization. The resistance decrease with an increase of temperature，$1.45\times10^6\Omega$ for 100K and $5.81\times10^5\Omega$ for 280K，exhibits a LGSCO phase separated behavior(two phase). Tc is determined from the R-1/T curve, which measured an IM transition at 128 K. The Tc goes up to a $<r_A>$ value of LGSCO and decease therefore. The resistance of cobaltates also increases with the effects of cation size and mismatch disorder. This result also show the effect of carriers captured at interface. Both $La_{0.4}Gd_{0.1}Sr_{0.5}CoO_3$ and ZnO show the expected metallic behavior, but orthorhombic $Gd_{0.5}Sr_{0.5}CoO_3$ show a slight departure from metallic behavior. While $La_{0.4}Gd_{0.1}Sr_{0.5}CoO_3$ at low temperature is metallic, $La_{0.4}Gd_{0.1}Sr_{0.5}CoO_3$ at high temperature is an insulator.

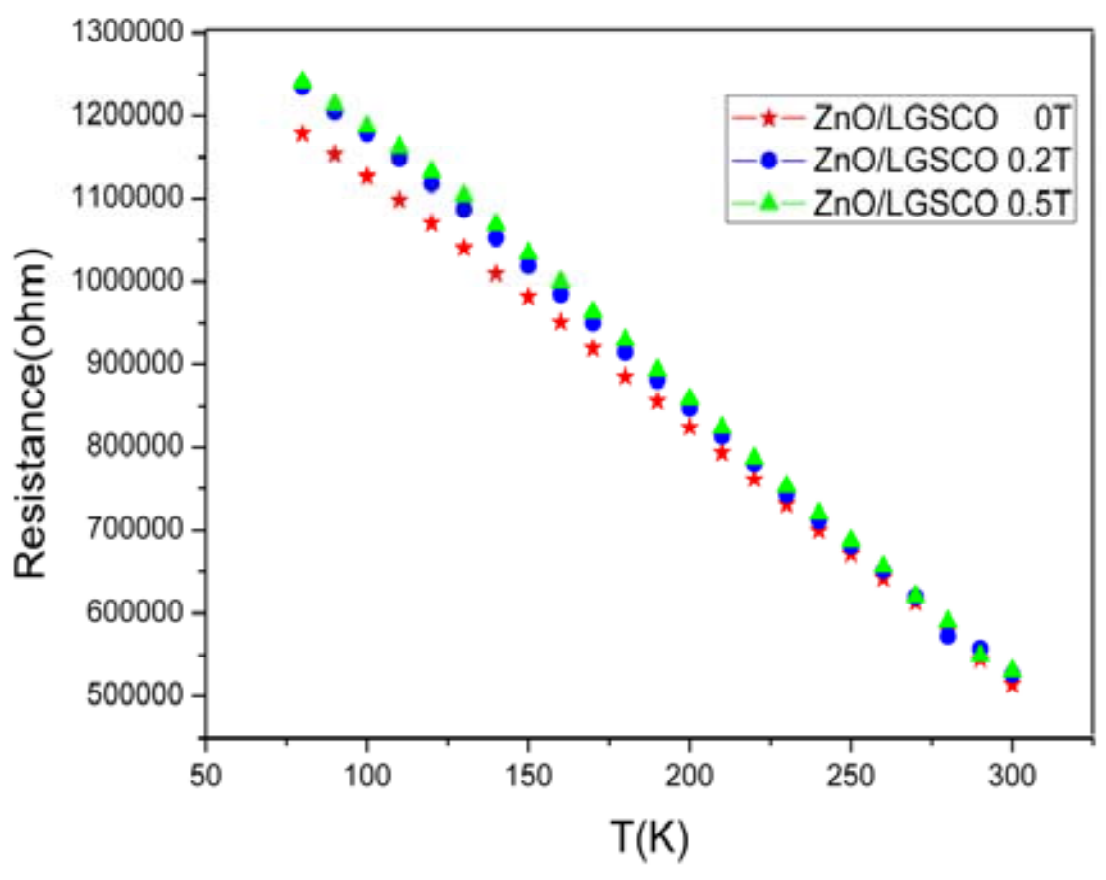

Fig4. Temperature dependence of LGSCO/ZnO junction resistance of the heterostructure under applied different magnetic field.

**(2) Magnetic feature of phase separated state**

The experiment shows important role of cation size and disorder on both the electric and the magnetic resistance properties. A large lattice mismatch led to more defects at interface, and captured more carriers. The lower the temperature，the more carriers will be captured, and the junction resistance is increased. On the other hand, the electrons transport and M-I transition based on double exchange and polarons hopping were due to spin-orbit coupling between localized electrons Co-3d $t_{2g}$ and itinerant electrons O-2p $e_g$ in the films. With an increase of temperature above Tco, the sample properties represent the intensity characteristics to the single glass phase and COI insulator, because the charge-orbital ordering transition. With a decrease of temperature below Tc shown as Fig5 (a) behaviour of Log ρ versus T for the LGSCO/ZnO (b) behaviour of ρ/T versus 1/T for the LGSCO/ZnO, the components represent the intensity properties to two phases, ferromagnetic and anti ferromagnetic ordering, because long c lattice constant is for spin transition. Such state is described to the strong coupling system between the orbital and lattice degree of freedom via the Jahn-Teller effect. The magnetoresistance is defined as $MR=\Delta R/R=(R_H-R_O)/Ro$ where $R_H$ is the resistance of heterojunction with the applied magnetic field, $R_O$ is the resistance without the magnetic field shown in Fig.5c. The temperature was dependence of junction $\Delta R/R$ and MR of the heterostructure. We could measure the charge-orbit variation and spin-orbital coupling from junction $\Delta R/R$ and MR. Here, we have exhibited lattice constant C for the magnetic feature: short c is for ferromagnetic and antiferromagnetic type, further short c for COI. As well as, we described the long c for ferromagnetic and

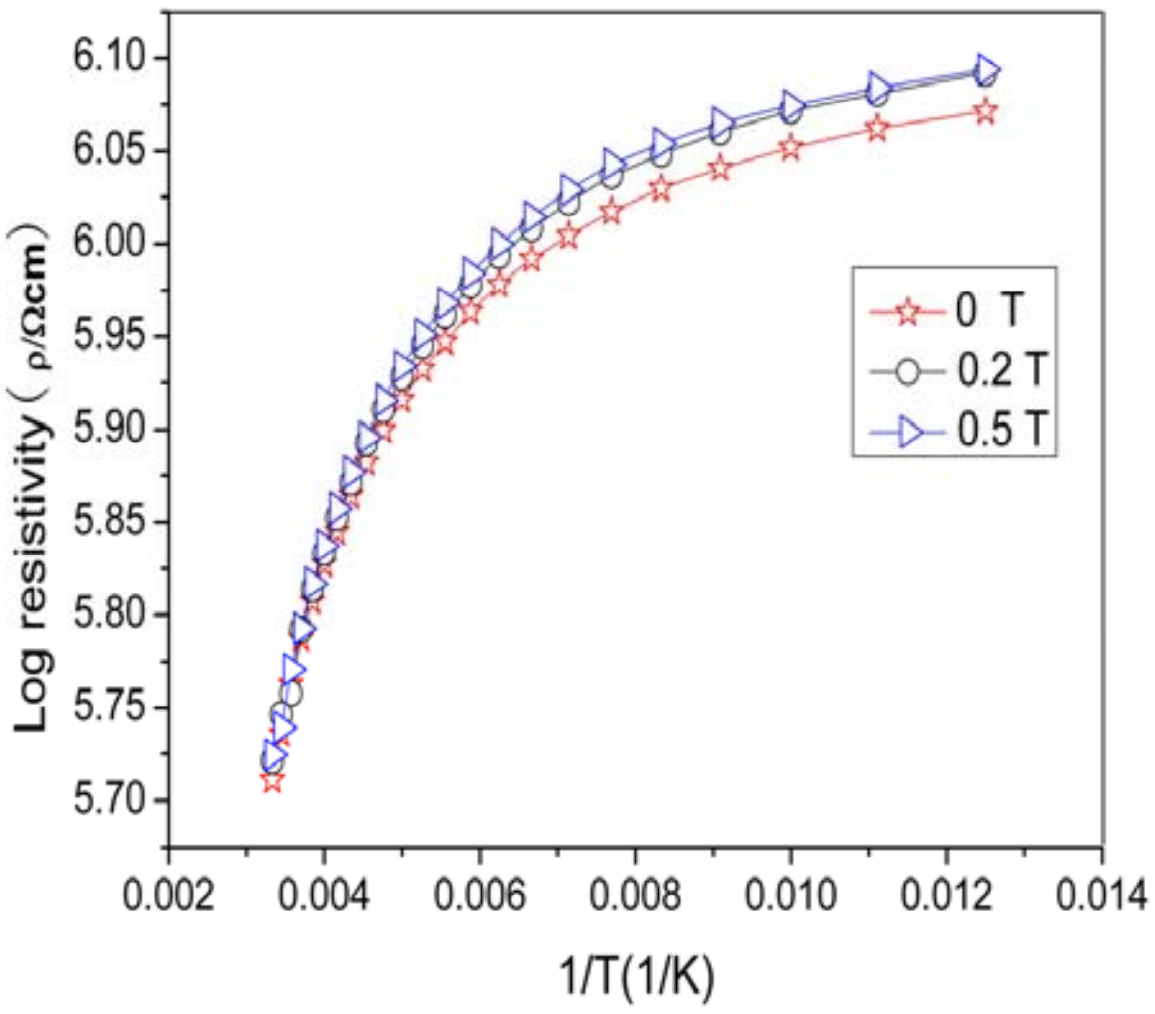

(a)

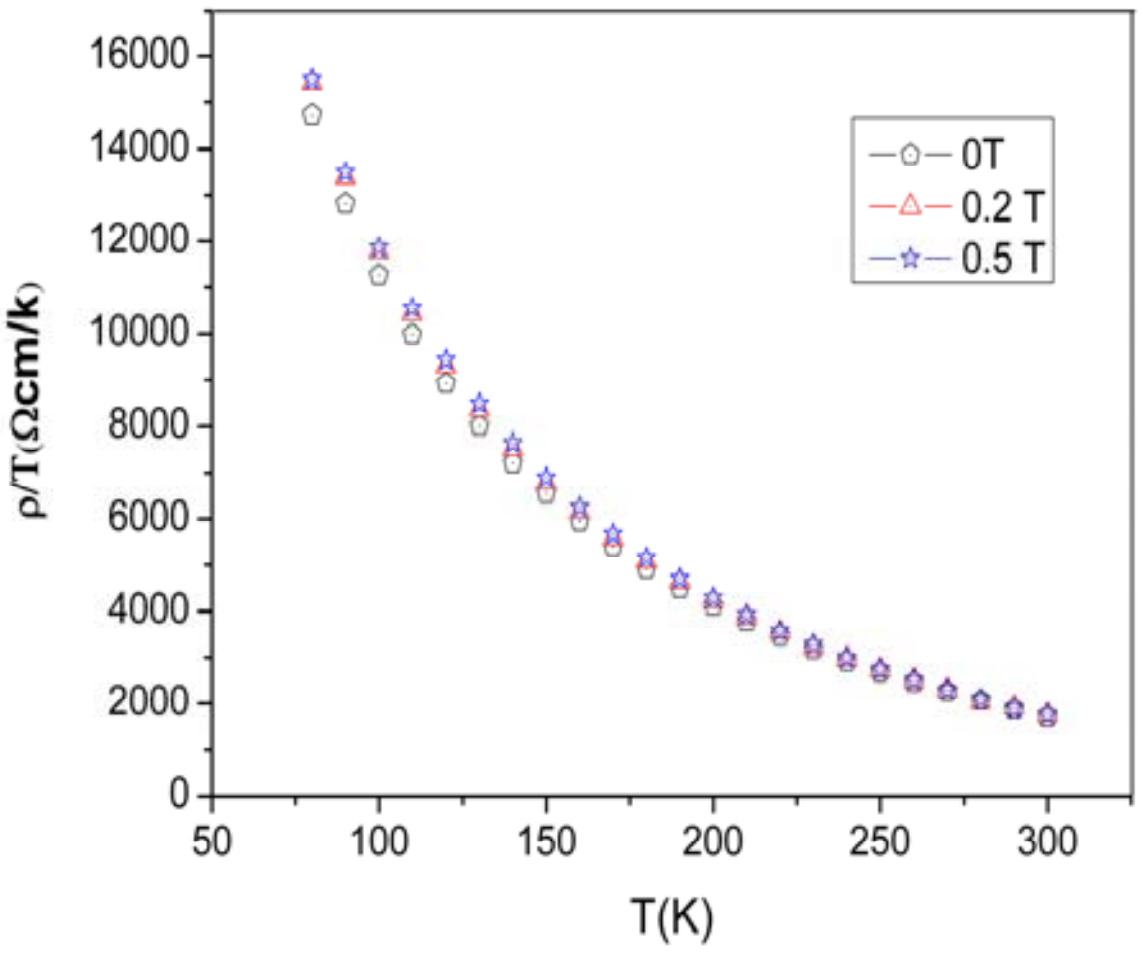

(b)

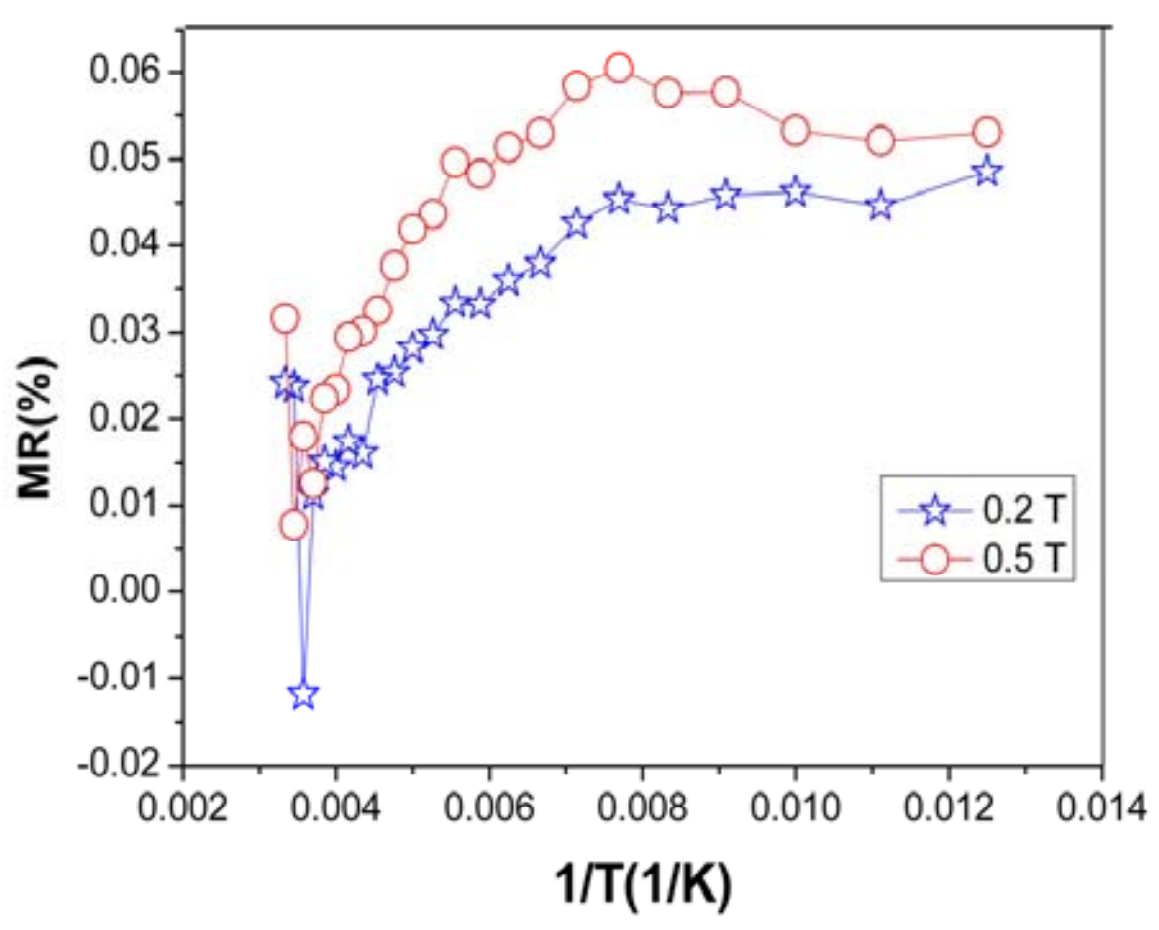

(c)

Fig5. (a) Behaviour of Log ρ versus T for the LGSCO/ZnO (b) Behaviour of ρ/T versus 1/T for the LGSCO/ZnO (c)Temperature dependence of the LGSCO/ZnO junction magnetic resistance at magnetic field H = 0.2 and 0.5 T.

antiferromagnetic spin. The LGSCO MR had MIT properties and exhibited a steep increase in the LGSCO single layer structure at the low temperature. Over the range of temperature less than 130 K, the junction displayed a metallic-like in which there was temperature dependence of d(MR)/dT>0. Coexisting of the ferromagnetic and antiferromagnetic ordering could be concluded to a canting spin structure, such the case of the double layer manganites $La_{2-2x}Sr_{1+2x}Mn_2O_7$. Actually, the ρ-T curve at y=0.5 Co shows a significant I-M behavior at around Tc=127 K.[1-=19,21-26]

Moreover, domain structure takes effect on the boundary resistivity of LGSCO film, finally varied the charge-ordering temperature and Curie temperature, respectively. COI above TCO stand for charge-ordering insulator, thus, PI and FM stand for the paramagnetic insulator and ferromagnetic metallic phase at $T_C$. domain wall orientations and domain structure strongly influence the paramagnetic insulator properties of LGSCO，furthermore，change the junction properties at LGSCO/ZnO thin film.

The phase field dynamics was adopted to explore possibilities of substrate effect and external field on the lattice. The anisotropy caused by changes of substrate stress and magnetic field, is enough to produce observation area of domain structure change. We adopted the two rectangular systems, one is local system which the pseudocubic crystal cell, other is globe system with orthogonal axes $x_i'$, $x_i'$ and $x_i'$ in the film plane.

We begin the simulation with small random polarization. The $P_3$ polarization component was built by the magnetic field and voltage resulting from the film/substrate interface. The evolution of phase field of polarization is controlled by the time-dependent Ginzburg-landau and Cahn-Hilliard free energy equation.[13,14], $\frac{\partial P_i(r,t)}{\partial t} = -L\frac{\delta F}{\delta P_i(r,t)} \quad (i=1,2,3)$, Where L is the kinetic coefficient, and the L is related with domain wall mobility. F is the total free energy including Landau, gradient, electrostatic and elastic energies.

Our observation clearly demonstrates that the magnetic domain structures in LGSCO vary significantly with the doping level. [7,8,24-27] The polarizations within these two doping level variants are related to $71^0$ and $109^0$ domain walls. The domain are oriented along (100) or (010) directions of the substrate. The red and yellow are typically paired.

We obtain the ferroelectronic phase image shown in Fig. 6, the domain splits into several parts. When the sample was warming up above Tc, and the sample was cooled from above Tc to lower temperature, the properties of domain appear. The domain stared to appear at medium temperature. The domain structure of polycrystalline LGSCO thin film of the spontaneous polarization process (The number of iterations step t=3000-5000) was shown in Fig 6.

Further, the phase transition of junction from ferromagnetic metal to paramagnetic insulator occurred at Curie temperature Tc 127 K. The maximum value of the positive MR at 0.2 T is 4.86%, at 0.5 T is 6.05% for approximately 140K. There were main factors which influenced the spin orientation of $Co^{3+}$ and $Co^{4+}$.[15,28,29,30] We speculated that the junction resistance was mostly derived from the interface diffusion of ZnO/LGSCO due to the lattice mismatch between ZnO and LGSCO which the junction resistance in ZnO and LGSCO films was altered by the ZnO/LGSCO interfacial phase cluster, bond length and the band angle of $Co^{3+}$-O-$Co^{4+}$ chain. Thus, the carrier transfer of heterjunction was derived from the domain boundary scattering and capture effect, which were controlled by optical and magnetic perturbations.

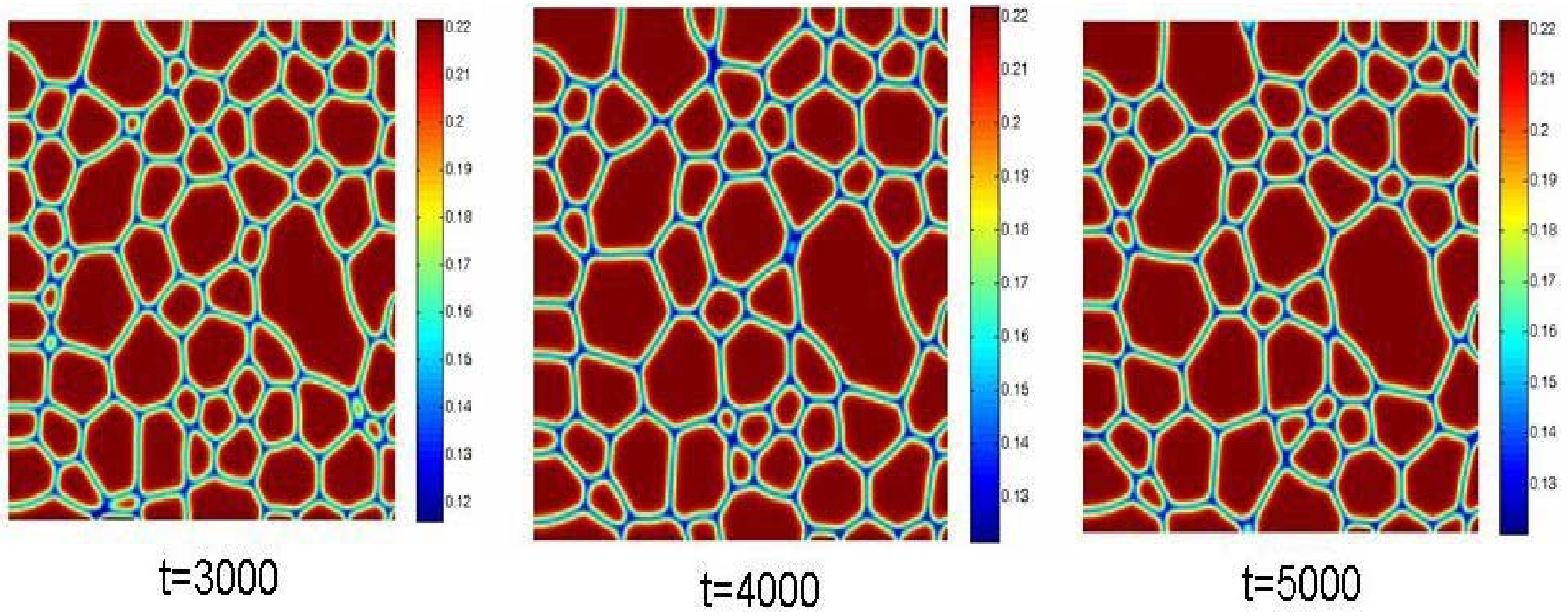

Fig.6. The domain structure of polycrystalline LGSCO thin film of the spontaneous polarization process (The number of iterations step t=3000-5000)

#### 4. CONCLUSIONS

In summary, the presence of FM phase and carrier injection effect mechanism is clearly demonstrated in the $ZnO/La_{0.4}Gd_{0.1}Sr_{0.5}CoO_3$. The ferromagnetic transformation into the phase-separated (two phase) state were observed below Tc~127 and have observed the lattice change discontinuously in the doped cobalt perovskites $La_{0.4}Gd_{0.1}Sr_{0.5}CoO_3$. We use Ginzburg-Landau-Devonshire phase-field model to observe the evolution of ferromagnetic and ferroelectric domain structure LGSCO thin film. It was found that the domain structures of thin film were strongly dependent on the combination of force electric coupling. The positive MR effect were observed 4.86% for 0.2 T, 6.05% fort 0.5 T in the $ZnO/La_{0.4}Gd_{0.1}Sr_{0.5}CoO_3$/LaAlO heterostructure at approximately 140K. The ZnO/ LGSCO junction exhibited the carrier transfer of MIT transition at different temperatures. The electric-mechanical coupling was introduced by domain anisotropy due to temperature, optical and magnetic perturbations. This inhomogeneous should play a crucial role in the the further understanding of charge order, spin-orbital coupling, magnetic order, optical field and lattice structure from the view of energy band in ZnO/ $La_{0.4}Gd_{0.1}Sr_{0.5}CoO_3$/LaAlO and positive CMR effect found in doped cobalt perovskites.

**ACKNOWLEDGEMENTS**

This work is supported by MOE Key Laboratory for Nonequilibrium of China 2009 (Grant No. 2009, 1001).